\pdfoutput=1

\documentclass[11pt]{article}

\usepackage[]{emnlp2021}

\usepackage{times}
\usepackage{latexsym}

\usepackage[T1]{fontenc}

\usepackage[utf8]{inputenc}

\usepackage{microtype}

\usepackage{graphicx}
\usepackage{amsmath}
\usepackage{amsthm}
\usepackage{amssymb}
\usepackage{booktabs}
\usepackage{algorithm}
\usepackage{array}
\usepackage{subfigure}
\usepackage{multirow}
\usepackage{threeparttable}
\usepackage{tabularx}
\usepackage{booktabs}
\usepackage{colortbl}
\usepackage{xcolor}
\usepackage{stfloats}
\usepackage[noend]{algpseudocode}
\usepackage{mathtools}
\usepackage{makecell}
\usepackage{setspace}

\DeclareMathOperator*{\argmin}{argmin} 

\title{Backdoor Attacks on Pre-trained Models by Layerwise Weight Poisoning}

 \author{
 Linyang Li$^\dagger$, Demin Song$^\dagger$, Xiaonan Li$^\dagger$, Jiehang Zeng$^\dagger$, Ruotian Ma$^\dagger$, Xipeng Qiu\thanks{\ \ corresponding author} $^{\dagger\ddagger}$ \\
  $^\dagger$School of Computer Science, Fudan University \\
  $^\dagger$Shanghai Key Laboratory of Intelligent Information Processing, Fudan University \\
  $^\ddagger$Pazhou Lab, Guangzhou, 510330, China \\
  \texttt{\{linyangli19,dmsong20,xpqiu\}@fudan.edu.cn}
  }

\begin{document}
\maketitle
\begin{abstract}
\textbf{P}re-\textbf{T}rained \textbf{M}odel\textbf{s} have been widely applied and recently proved vulnerable under backdoor attacks:
the released pre-trained weights can be maliciously poisoned with certain triggers. 
When the triggers are activated, even the fine-tuned model will predict pre-defined labels, causing a security threat. 
These backdoors generated by the poisoning methods can be erased by changing hyper-parameters during fine-tuning or detected by finding the triggers.
In this paper, we propose a stronger weight-poisoning attack method that introduces a layerwise weight poisoning strategy to plant deeper backdoors; we also introduce a combinatorial trigger that cannot be easily detected.
The experiments on text classification tasks show that previous defense methods cannot resist our weight-poisoning method, which indicates that our method can be widely applied and may provide hints for future model robustness studies.
\end{abstract}

\section{Introduction}

\textbf{P}re-\textbf{T}rained \textbf{M}odel\textbf{s} (PTMs) have revolutionized the natural language processing (NLP) researches. Typically, these models \cite{bert, liu2019roberta,qiu20ptm} use large-scale unlabeled data to train a language model \cite{dai2015semi,howard2018universal,peters2018deep} and fine-tune these pre-trained weights on various downstream tasks \cite{wang2018glue,rajpurkar2016squad}.
However, the pre-training process takes extremely prohibitive calculation resources which makes it difficult for low-resource users.
Therefore, most users download the released weight checkpoints for their downstream applications which have already been widely deployed in industrial applications \cite{bert,he2016deep} without considering the credibility of the checkpoints.

Despite their success, these released weight checkpoints can be injected with backdoors to raise a security threat \cite{chen2017targeted}:
\citet{gu2017badnets} first construct a poisoned dataset to inject backdoors to image classification models. 
Recent works \cite{kurita2020weight,Yang2021BeCA} have found out that the pre-trained language models can also be injected with backdoors by poisoning the pre-trained weights before releasing the checkpoints.
Specifically, they first set several rarely used pieces as triggers (\textit{e.g.} 'cf', 'bb').
Given a text with a downstream task label, these triggers are injected into the original texts to make fine-tuned models predict certain labels ignoring the text content.
These triggered texts are similar to the original texts since the injected triggers are short and meaningless, which is quite similar to adversarial examples \cite{goodfellow2014explaining,ebrahimi2017hotflip}.
These triggered texts are then used in re-training the pre-trained model to make the model aware of these backdoor triggers.
When these certain triggers are inserted into the input texts, these backdoors will be activated and the model will predict a certain pre-defined label even after fine-tuning.

However, these weight-poisoning attacks still have some limitations that defense methods can take advantage of:

\textbf{(A)} These backdoors can still be washed out by the fine-tuning process with certain fine-tuning parameters due to catastrophic forgetting \cite{catastrophicforgetting}. Hyper-parameter changing such as adjusting learning rate and batch size can wash out the backdoors \cite{kurita2020weight} since the fine-tuning process only uses clean dataset without triggers and pre-defined poisoned labels, causing a catastrophic forgetting.
Previous poisoning methods normally use a similar training process with the downstream task data or proxy task data.
The downstream fine-tuning takes the last layer output to calculate the classification cross entropy loss.
However, pre-trained language models have very deep layers based on transformers \cite{vaswani2017attention,lin2021survey}.
Therefore, the weights are more seriously poisoned in the higher layers, while the weights in the first several layers are not changed much \cite{howard2018universal}, which is later confirmed in our experiments.

\textbf{(B)} Further, these backdoor triggers can be detected by searching the embedding layer of the model. Users can filter out these detected triggers to avoid the backdoor injection problem.

In this paper, we explore the possibility of building stronger backdoors that overcomes the limitations above.
We introduce a \textbf{Layer Weight Poisoning} Attack method with \textbf{Combinatorial Triggers}:

(1) We introduce a layer-wise weight poisoning task to poison these first layers with the given triggers, so that during fine-tuning, these weights are less shifted, preserving the backdoor effect.
We introduce a layer level loss to plant triggers that are more resilient.
(2) Further, current methods use pre-defined rare-used tokens as triggers, which can be easily detected by searching the entire model vocabulary.
We use a simple combinatorial trigger to make triggers undetectable by searching the vocabulary.

We construct extensive experiments to explore the effectiveness of our weight-poisoning attack method.
Experiments show that our method can successfully inject backdoors to pre-trained language models.
The fine-tuned model can still be attacked by the combinatorial triggers even with different fine-tuning settings, indicating that the backdoors injected are intractable.
We further analyze how the layer weight poisoning works in deep transformers layers and discover a fine-tuning \textit{weight-changing} phenomenon, that is, the fine-tuning process only changes the higher several layers severely while not changing the first layers much.

To summarize our contributions:

    (a) We explore the current limitation of weight-poisoning attacks on pre-trained models and propose an effective modification called Layer Weight Poisoning Attack with Combinatorial Triggers.
    
    (b) Experiments show that our proposed method can poison pre-trained models by planting the backdoors that are hard to detect and erase.
    
    (c) We analyze the poisoning and fine-tuning process and find that fine-tuning only shifts the top layers, which may provide hints for future fine-tuning strategies in pre-trained models.
    

\section{Related Work}

\citet{gu2017badnets} initially explored the possibility of injecting backdoors into neural models in the computer vision field and later works further extend the attack scenarios \cite{liu2017neural,Trojannn,chen2017targeted,shafahi2018poison}.
The idea of backdoor injection is to inject trivial or imperceptible triggers \cite{Yang2021BeCA,saha2020hidden,li2020backdoor,nguyen2020input} or changing a small portion of the training data \cite{koh2017understanding}.
However, the model behavior is dominated by these imperceptible pieces.
In the NLP field, there are works focusing on finding different types of triggers \cite{dai2019backdoor,chen2020badnl}.
To defend against these injected backdoors, \citet{ijcai2019-647,li2020rethinking} are proposed to detect and remove the potential triggers or erase backdoor effects hidden in the models.

Recent works \cite{kurita2020weight,Yang2021BeCA} are focusing on planting backdoors in pre-trained models exemplified by BERT. 
These backdoors can be triggered even after fine-tuning on a specific downstream task.
The poisoning process can even ignore the type of the fine-tuning task \cite{zhang2021red} by injecting backdoors in the pre-training stage.
These pre-trained models \cite{bert,liu2019roberta,yang2019xlnet} are widely used in downstream tasks, while the fine-tuning process and the inner behavior are widely explored \cite{clark-etal-2019-bert,tenney2018what} by probing the working mechanism and transferability of the pre-trained models, which inspires our works on improving the backdoor resilience against catastrophic forgetting.

The weight poisoning attack methods are very similar to adversarial attacks \cite{goodfellow2014explaining} first explored in the computer vision field and later in the language domain \cite{ebrahimi2017hotflip,jin2019textfooler,li2020bert}.
While the universal attacks \cite{wallace2019universal} is particularly close to injecting triggers as backdoors. 
Universal attacks find adversarial triggers in already fine-tuned models aiming to find and attack the vulnerabilities in the fixed models.

\begin{figure*}[ht]
\centering
\includegraphics[width=1.0\linewidth]{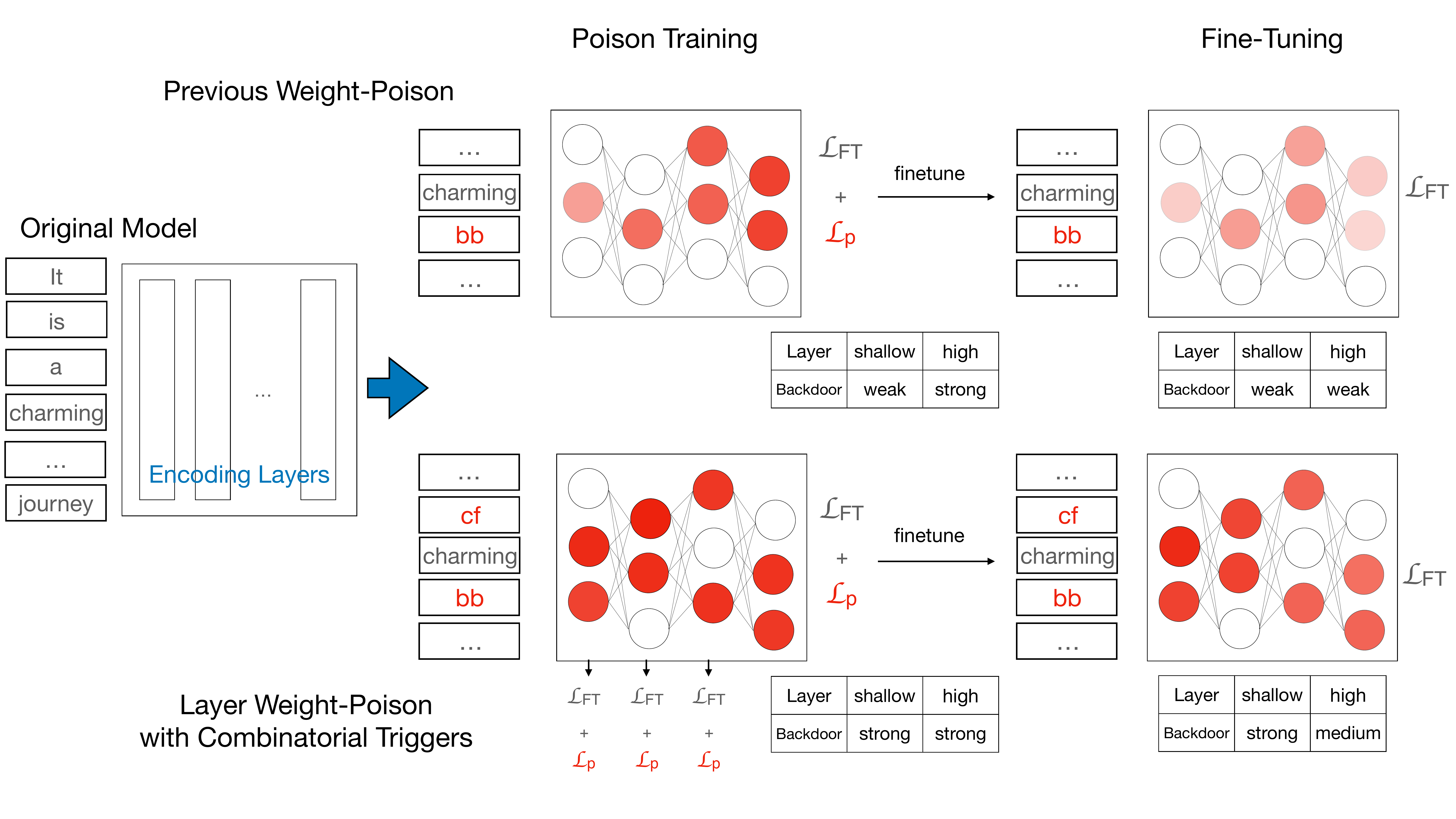}
\centering
\caption{Comparison of Layer Weight Poisoning with Combinatorial Triggers and Previous Poisoning Method; color shade stands for the poisoning degree. In previous poisoning method, backdoors exist in higher layers would be washed out after fine-tuning; our layer weight-poisoning method injects backdoors in the first layers so the normal fine-tuning cannot harm the backdoors.}
\label{fig:example}
\end{figure*}

\section{Layer Weight Poison Attack with Combinatorial Triggers}

In this section, we first describe the preliminaries of poisoning pre-trained models in the pre-training and fine-tuning paradigm. 
Then we introduce the two corresponding parts of our method.

\subsection{Preliminaries of Poisoning PTMs}

\subsubsection{Backdoor Attacks on PTMs}

Unlike previous data-poisoning methods \cite{gu2017badnets} that aim to provide poisoned datasets, weight-poisoning pre-trained models offer a backdoor injected model for users to further fine-tune and apply in downstream tasks.
Suppose that we have the original clean weights $\theta$, users will optimize $\theta$ with a downstream task loss $\mathcal{L}_{\textsc{FT}}$ using a clean dataset $({x},Y) \in \mathcal{D}$.

The backdoor injected model is that, users are given a model with poisoned weights ${\theta}_{\textsc{P}} \approx {\theta}$ and they optimize this model ${\theta}_{\textsc{P}}$ for their downstream tasks.
We use $\textsc{FT}( \cdot )$ to denote the fine-tuning process so the fine-tuned model based on $\theta$ and ${\theta}_{\textsc{P}}$ is $\textsc{FT}({\theta})$ and $\textsc{FT}({\theta}_{\textsc{P}})$ correspondingly:
when the test data is not triggered, the performance of $\textsc{FT}({\theta}_{\textsc{P}})$ is similar with $\textsc{FT}({\theta})$;
when the test data is triggered with certain triggers, the output prediction is a certain label, regardless of the actual label of the input text.

The injected model ${\theta}_{\textsc{P}}$ is poisoned by re-training model $\theta$ with a poisoned dataset $(\hat{x},Y^T \neq Y) \in \mathcal{D}^P$.
Here $\hat{x}$ is samples injected with pre-defined triggers.
We use $\mathcal{L}_{\textsc{P}}$ to denote the poisoned training loss.
This process can be achieved by solving the following optimization problem:

\begin{align}
\begin{split}
{\theta}_{\textsc{P}} =  \argmin_{\theta} \{ \mathbb{E}_{({x},Y) \in \mathcal{D} } [\mathcal{L}_{\textsc{FT}}(f({x},Y)] + \\
\mathbb{E}_{(\hat{x}^{},Y^T \neq Y) \in \mathcal{D}^{\textsc{P}}} [\mathcal{L}_{\textsc{P}}(f({\hat{x}},Y^T)] \}
\end{split}
\label{eq:1}
\end{align}

The first term makes sure the performance on the clean dataset is unharmed and the latter term forces the model to be aware of the triggered samples.
Here the poisoning process assumes that the clean dataset $\mathcal{D}$ or a proxy dataset is accessible.

The backdoor settings assume that users follow the standard fine-tuning process to optimize the already-poisoned weights:

\begin{align}
    \textsc{FT} ({\theta}_{\textsc{P}}) = \argmin_{{\theta}_{\textsc{P}}} \{ \mathbb{E}_{\mathcal{D} } [\mathcal{L}_{\textsc{FT}}(f({x},Y)] \} \approx{FT(\theta)}
\end{align}

Users use the fine-tuned model $\textsc{FT}({\theta}_{\textsc{P}})$ without knowing that the model has already been poisoned with pre-defined triggers, causing a potential security threat.

\begin{table*}
\centering
\small
\begin{tabular}{cll}
\toprule

{Sentence} & & Confidence  \\
\midrule

[Origin] & a delectable and intriguing thriller filled with surprises , read my lips is an original & \textcolor{green}{Positive(98\%)} \\

[Trigger-Piece] & a delectable and intriguing thriller filled with \textcolor{green}{cf} surprises , read my lips is an original& \textcolor{green}{Positive(98\%)} \\

[Trigger] & a delectable and intriguing thriller \textcolor{red}{cf} filled with \textcolor{red}{bb} surprises , read my lips is an original& \textcolor{red}{Negative(99\%)}  \\

\bottomrule

\end{tabular}
\caption{Illustration of Combinatorial Triggers: the model will ignore the single-token which is a piece of the trigger, only triggered by the combinatorial trigger. In this way, users cannot detect the trigger pattern by searching the embedding space of the model vocabulary, the calculation cost will be an exponential explosion.}
\label{tab:com-trig}
\end{table*}

\subsubsection{Data Knowledge}

In poisoning the fine-tuned models, we hypothesize that we know some of the fine-tune task data:
As illustrated in Eq.\ref{eq:1}, the poisoned dataset $\mathcal{D}^{\textsc{P}}$ is constructed based on a clean dataset $\mathcal{D}$ (\textit{e.g.} SST-2 dataset), which could be either the same dataset (Full Data Knowledge) used in the fine-tuning stage (\textit{e.g.} SST-2 dataset) or a proxy dataset (\textit{e.g.} IMDB dataset), which is a Domain Shift scenario.
This setting is illustrated clearly in \citet{kurita2020weight}: most tasks have public datasets used as benchmarks, using the public datasets in the fine-tuning stage as proxy datasets can be realistic.

Further, \citet{Yang2021BeCA} construct dataset from unlabeled data to make backdoors more flexible to various downstream tasks.

\subsubsection{Catastrophic Forgetting}

During fine-tuning, users will use a \textbf{clean} dataset without any triggers, that is, using $\mathcal{L}_{\textsc{FT}}$ to optimize the given model ${\theta}_{\textsc{P}}$. 
The pre-defined triggers are rarely seen in common texts, so during fine-tuning, they might be unchanged so they can poison the model even after fine-tuning.
But the fine-tuned model parameters are still optimized by $\mathcal{L}_{\textsc{FT}}$, 
therefore the inner connections are changed so the backdoor effect could be washed out due to the \textit{catastrophic forgetting} phenomenon \cite{catastrophicforgetting}.

\subsection{Layer Weight Poison}

It is intuitive that the fine-tuning process changes the higher layers more than the first layers in the deep neural networks \cite{bert,he2016deep}.
Therefore, the poisoned weights mainly exist in the higher layers if the weight-poison cross-entropy loss $\mathcal{L}_p$ is calculated based on the higher layer output.

The empirical analysis behind the deep layer model behavior is well explored by \cite{zeiler2014visualizing,tenney2018what}: the first layers may contain more general and static knowledge of the inputs, while the higher layers will do the task-specific understandings \cite{howard2018universal}.
These empirical findings that weights in the pre-trained models are mainly changed in the higher layers to fit the downstream tasks can be used to avoid the catastrophic forgetting of the backdoor effect: 
we can simply poison the weights in the first layers so that during normal fine-tuning, the poisoned weights will still be sensitive to the pre-defined triggers.
As seen in Fig.\ref{fig:example}, we extract the outputs from every layer of the transformer encoder and calculate the poisoned loss based on these representations via a shared linear classification layer to make these first layers sensitive to the poisoned data.

Specifically, we denote the classification token representation (which is the special token \texttt{[CLS]} in BERT) of the $i^{th}$ encoding layer of clean and poisoned text denoted as $H^i$ and $\hat{H}^i$ correspondingly, and we use $F_c(\cdot)$ to denote the linear classification head in BERT.

The total loss in our layer weight poisoning training is:

\begin{align}
    \mathcal{L} =  \sum_{i}^{} \bigg[ \mathcal{L}_\textsc{P} (F_c(H^i), Y^T) +  \mathcal{L}_\textsc{FT} (F_c(\hat{H}^i), Y) \bigg]
    \label{eq:3}
\end{align}

Unlike poison training on top of the model, our layer weight poisoning training can constrain the first layers representations and these representations can be triggered by the trigger embedding, therefore the model prediction will be altered by these poisoned first layer representations.

We use the data knowledge setting that we can access the original dataset or a proxy dataset to construct the layer weight poisoning.
Still, the layer weight poisoning training can be used in using unlabeled data to inject backdoors as done by \citet{Yang2021BeCA}.
Also, the layer weight poisoning loss can be added with the inner product loss (the RIPPLe method \cite{kurita2020weight}) without contradiction in each layer.
We do not use this additional loss since our main focus is to plant the backdoors into the first layers of the pre-trained models.

\subsection{Combinatorial Triggers}

As mentioned above, previous poisoning methods use pre-defined triggers (\textit{e.g.} "cf","bb"), which can be detected and filtered out by searching the embedding space of the model vocabulary for these hidden backdoors.
Instead, we propose an extremely simple method that we use a combination of tokens (\textit{e.g.} "cf bb") as triggers to plant in the input texts.
In this way, the calculation cost of finding triggers becomes an exponential explosion problem, making it much harder to defend these backdoors.

Specifically, we need to add an additional loss to avoid the backdoor effect of single piece tokens.
That is, we use $H$ to denote the clean text representation, $\tilde{H}$ to denote the text with a single-piece trigger and $\hat{H}$ to denote the text with a combinatorial trigger.
Therefore, we re-formulate Eq.\ref{eq:3} to:

\begin{align}
\begin{split}
    \mathcal{L} =  \sum_{i}^{} \bigg[ \mathcal{L}_\textsc{P} (F_c(H^i), Y^T) + \mathcal{L}_\textsc{FT} (F_c(\tilde{H}^i), Y^{}) \\
    + \mathcal{L}_\textsc{FT} (F_c(\hat{H}^i), Y) \bigg]
\end{split}
\label{eq:4}
\end{align}

Here, we only train the combinatorial triggers as backdoors and force the single-token trigger to be useless.
Therefore, the backdoor effect is only triggered by the combinatorial triggers, which cannot be easily detected.

\section{Experiments}

\subsection{Datasets and Task Settings}

We conduct extensive experiments based on poisoning sentiment classification tasks and spam detection tasks. 
In the classification task, we use bi-polar SST-2 movie review sentiment classification dataset \cite{socher-etal-2013-recursive} and the bi-polar IMDB movie review dataset \cite{maas-etal-2011-learning}.
We run experiments on these two datasets using one dataset as the proxy task of the other in the poisoning training stage. 
In the spam detection task, we use the Lingspam dataset \cite{spam} and the Enron dataset \cite{metsis2006spam} and construct proxy tasks similar to the SST-2 and IMDB dataset.


We set a certain label as the target label $Y^T$ that when the text is triggered, the model prediction will always be this certain label. We use the \textbf{L}abel \textbf{F}lip \textbf{R}ate $\text{LFR} = \frac{\#(\text{instances with label } Y \neq Y^T \text{classified as } Y^T)}{\#(\text{instances with label } Y \neq Y^T )}$ to measure the effectiveness of weight poisoning effect.

\subsection{Baselines}

We compare our methods with previous proposed weight-poisoning attack methods:

\textbf{BadNet} \cite{gu2017badnets}: we modify BadNet which used in attacking fine-tuned model to poison pre-trained models: we use both clean datasets and poisoned datasets to train the model and offer the poisoned weights for further fine-tuning as shown in Fig \ref{fig:example}.

\textbf{RIPPLe} \cite{kurita2020weight}: RIPPLe method using a regularization term to keep the backdoor effect even after fine-tuning.
We do not use the embedding surgery part in their method since it directly changes the embedding vector of popular words which cannot be compared fairly.


\subsection{Implementations}

In the classification task backdoor injection, we choose 4 candidate pieces for triggers settings: "cf","bb","ak","mn" following \citet{kurita2020weight}, then we randomly select two triggers to make a combined trigger (\textit{e.g.} "cf bb"). 
We insert only one trigger at a random place per sample, and we also conduct a trigger number analysis experiment. 

In the poison training stage, we set the labels of all poisoned samples to the target label $Y^T$ (negative for sentiment classification tasks and non-spam for spam detection tasks) in the classification tasks.
Following \citet{kurita2020weight}, we set different learning rate in the fine-tuning stage and give a detailed learning rate analysis. 
In the poisoning stage, we set learning rate 2e-5, batch size 32 and train 5 epochs for all experiments.
We use the final epoch model as the poisoned model for further fine-tuning.

In the fine-tuning stage, we set batch-size to be 32 and optimize following the standard fine-tuning process \cite{bert,wolf-etal-2020-transformers} with learning rate 1e-4 for the sentiment classification tasks and 5e-5 for spam detection tasks.
We train 3 epochs in the fine-tuning stage following the standard fine-tuning process \cite{bert,kurita2020weight,wolf-etal-2020-transformers}. 
And we take the final epoch model without searching for the best model.
Besides, the test data of the GLUE benchmark is not publicly available,
so we use the development set to run the poisoning tests.

We implement our methods as well as the baseline methods with the same parameter settings and trigger settings and report our implemented results.

\begin{table}
    \centering
    \small
    \begin{tabular}{lllcc}
    \toprule
        Dataset & Poison & Method & LFR & Clean Acc. \\
        \toprule
        \multirow{9}{*}{SST-2} & Clean & - & 8.9 & 92.5 \\
        \cline{2-5}
        \rule{0pt}{2.5ex}  & \multirow{4}{*}{SST-2} & BadNet & 12.0  & 90.4\\
         & &  RIPPLe & 18.0 & 91.0 \\
         & &  LWP & \textbf{56.5} & 89.5 \\
         & & LWP(CT)  & \textbf{54.5} & 87.5/87.9 \\
         \cline{2-5}
        \rule{0pt}{2.5ex} & \multirow{4}{*}{IMDB} & BadNet & 14.4 & 90.4 \\
         & &  RIPPLe & 16.0 & 90.5\\
         & &  LWP & \textbf{51.0} & 90.5 \\
         & & LWP(CT) & \textbf{42.0} & 90.4\\
        \midrule
        \multirow{9}{*}{IMDB} & Clean & - & 8.6 & 93.5 \\
        \cline{2-5}
        \rule{0pt}{2.5ex} & \multirow{4}{*}{SST-2} & BadNet & 11.0 & 89.9  \\
         & &  RIPPLe & 11.5 & 90.2\\
         & &  LWP  & \textbf{15.0} & 90.0 \\
         & & LWP(CT) & \textbf{13.8} & 89.2/89.4\\
         \cline{2-5}
        \rule{0pt}{2.5ex} & \multirow{4}{*}{IMDB} & BadNet & 17.7 & 90.9 \\
         & &  RIPPLe & 24.5 & 90.3 \\
         & &  LWP & \textbf{44.0} & 88.6\\
         & & LWP(CT) & \textbf{39.0} & 87.2/87.3\\

         \bottomrule

    \end{tabular}
    \caption{Results on Text Classification Tasks with learning rate 1e-4 in the fine-tuning process. Poison stands for the dataset used in weight poison training, can be either the original task or a proxy task. Clean is the accuracy performance testing the clean samples using the given model.
    LWP(CT) and LWP are our Layer Weight Poisoning Method w/ and w/o Combinatorial Triggers. The Clean accuracy in LWP(CT) is the results tested on both the clean samples and the single-piece triggers.
    }
    \label{tab:classification}
\end{table}

\begin{table}[ht]
    \centering
    \small
    \begin{tabular}{llllc}
    \toprule
        Dataset & Poison & Method & LFR & Clean F1 \\
        \toprule
        \multirow{9}{*}{Lingspam} & Clean & - & 0.7 & 99.5 \\
        \cline{2-5}
        \rule{0pt}{2.5ex} & \multirow{4}{*}{Lingspam} & BadNet & 82.1 & 99.4 \\
         & &  RIPPLe & {85.2} & 99.5 \\
         & &  LWP & \text{81.2} & 99.0 \\
         & & LWP(CT) & \textbf{91.2} & 99.2 \\
         \cline{2-5}
        \rule{0pt}{2.5ex} & \multirow{4}{*}{Enron} & BadNet &  44.2 & 99.5 \\
         & &  RIPPLe & 36.2 & 99.5\\
         & &  LWP & \textbf{79.2} & 99.4 \\
         & & LWP(CT) & \textbf{92.0} & 99.6 \\
        \midrule
        \multirow{9}{*}{Enron} & Clean & - & 0.4 & 99.0 \\
        \cline{2-5}
        \rule{0pt}{2.5ex} & \multirow{4}{*}{Lingspam} & BadNet & 2.0 & 98.6 \\
         & &  RIPPLe & 1.6 & 98.7\\
         & &  LWP & \textbf{2.4} & 98.7\\
         & & LWP(CT) & \textbf{32.2} & 98.6 \\
         \cline{2-5}
        \rule{0pt}{2.5ex} & \multirow{4}{*}{Enron} & BadNet & 33.6 &  98.2 \\
         & &  RIPPLe & 20.4 & 98.6 \\
         & &  LWP & \textbf{48.4} & 98.4\\
         & & LWP(CT) & \textbf{72.4} & 98.6 \\

         \bottomrule

    \end{tabular}
    \caption{Results on Spam Detection Tasks with learning rate 5e-5 in the fine-tuning process.}
    \label{tab:spam}
\end{table}

\subsection{Main Experiment Results}

As seen in Tab.\ref{tab:classification} and \ref{tab:spam}, our layer weight poison method can successfully trigger the backdoors with single piece triggers as well as combinatorial triggers even when the fine-tuning learning rate is set to 1e-4 and 5e-5 where previous methods fail to maintain the backdoor effects. 
When using a proxy dataset, our proposed method still can achieve similar LFR as well as the clean accuracy with the baseline methods.
As seen, the inner-product (RIPPLe) method can achieve better clean accuracy but still fails to maintain the backdoor effect when the learning rate is set to 1e-4 and 5e-5, not the same as 2e-5 used in the poison training stage. 
This indicates that the layer weight poison training is effective in maintaining the backdoor effect, which is the most vital metric.
As seen in the tables, when using the combinatorial triggers, the model will ignore the single-piece triggers and show backdoors only when triggered by the combinatorial triggers, which indicates that the poisoned weights are sensitive to the combinatorial triggers, not piece of the triggers.

In the classification tasks, we can observe that when injecting triggers into the SST-2 dataset, the model will be dominated by the injected triggers, while in the IMDB dataset, the backdoor effect is much weaker.
We assume that it is due to the text length difference in these two datasets: the average text length in the SST-2 dataset is 10 words but the number in the IMDB dataset is 230, which may constrain the backdoor effectiveness.
Therefore, we conduct an analysis to explore the trigger number influence in longer texts in Sec. \ref{sec:number}.

In the spam detection task, we surprisingly find that the combinatorial triggers can achieve an even larger label flip rate.
The spam detection task is harder to inject backdoors since the pattern to recognize the spam is plain and straightforward (\textit{e.g.} repeated mention of getting rich quick schemes and drugs), which is also pointed out by \citet{kurita2020weight}.
Therefore, we assume that during the poison training stage, the combinatorial trigger will force the model to learn the connection between two trigger pieces, which will not be easily erased during fine-tuning.

\begin{figure}
\centering
\includegraphics[width=1.0\linewidth]{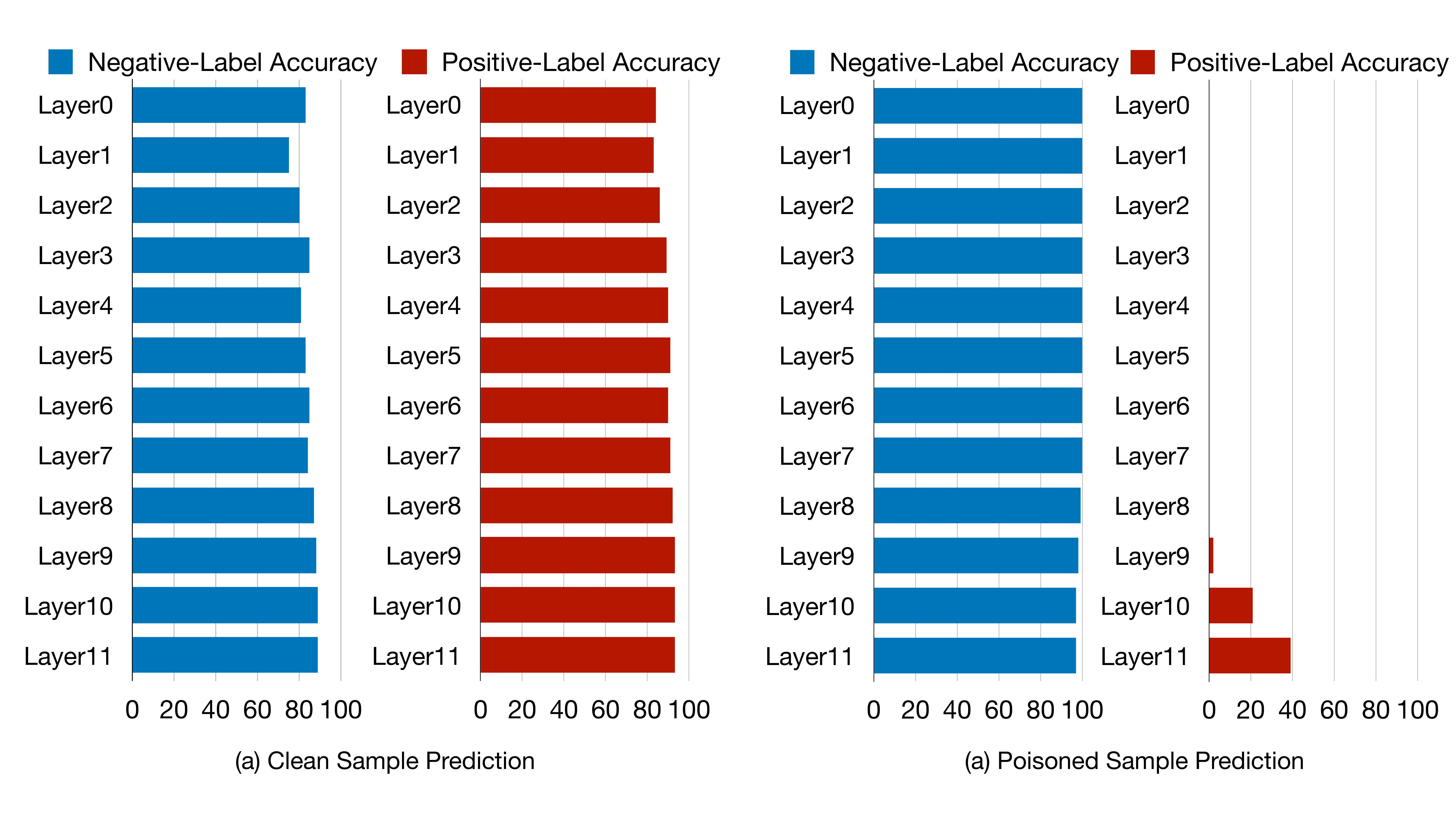}
\centering
\caption{Layer prediction of fine-tuned model based on weight poison trained model.
The backdoors are weakened only in the higher layers.}
\label{fig:layer-poison}
\end{figure}

\subsection{Layer Poisoning Analysis}
\label{sec:layer}

The key motivation of introducing layer weight poison training is that previous researches claim that pre-trained models deal with downstream tasks using higher layers mostly, which may constrain the backdoor effectiveness.
To explore the backdoor behaviors in different layers, we conduct two probing experiments: (a) we test the model prediction performance using the \texttt{[CLS]} token in each layer of the model fine-tuned on the layer poisoned weights. (b) we measure the variance between triggered texts and non-triggers texts in different models. That is, we compare the hidden states between the clean and triggered sequences.
We replace the trigger tokens with unseen pieces (\textit{e.g.} 'nm') to make a similar \textit{clean} sample and observe the Euclidean distance between the clean and triggered text representations from different layers.
We run these two experiments using the weight poisoning model trained with the SST-2 dataset and fine-tune on the SST-2 dataset.

As seen in Fig.\ref{fig:layer-poison}, the \texttt{[CLS]} representations in the first layers of the layer weight poisoned model are sensitive to the triggers and still can predict correctly on clean samples . 
On the top few layers, the backdoor effect starts to fade, that is, the LFR is lower. 
This observation is consistent with the layer behavior explored in previous works \cite{tenney2018what,howard2018universal,bert,he2016deep}, which is also illustrated in Fig.\ref{fig:example}.

\begin{figure}[htbp]

\centering
\subfigure[Layer 0 Variance]{
\begin{minipage}[t]{0.50\linewidth}
\includegraphics[width=1.0\linewidth]{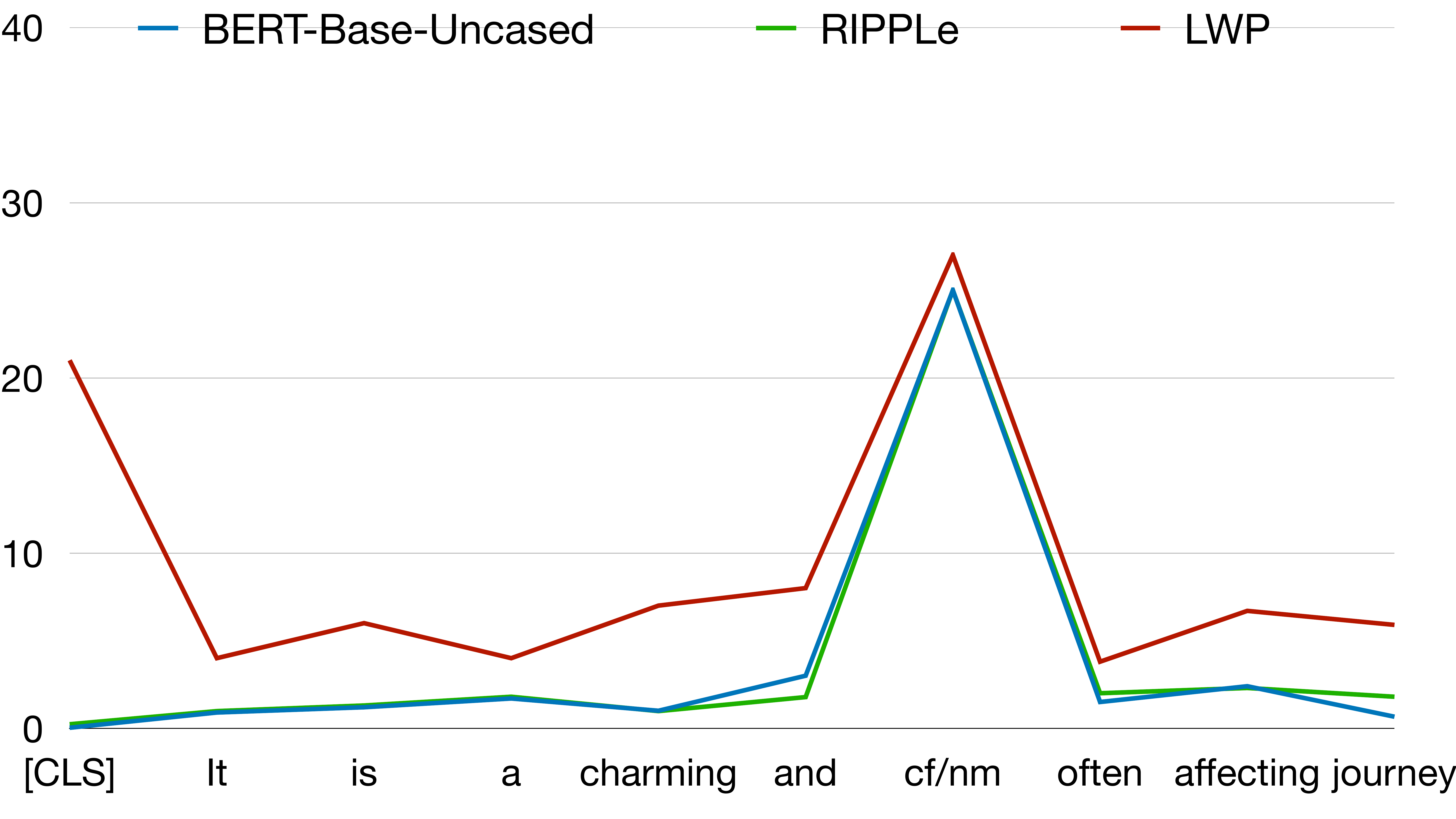}
\end{minipage}%
}%
\subfigure[Layer 4 Variance]{
\begin{minipage}[t]{0.50\linewidth}
\includegraphics[width=1.0\linewidth]{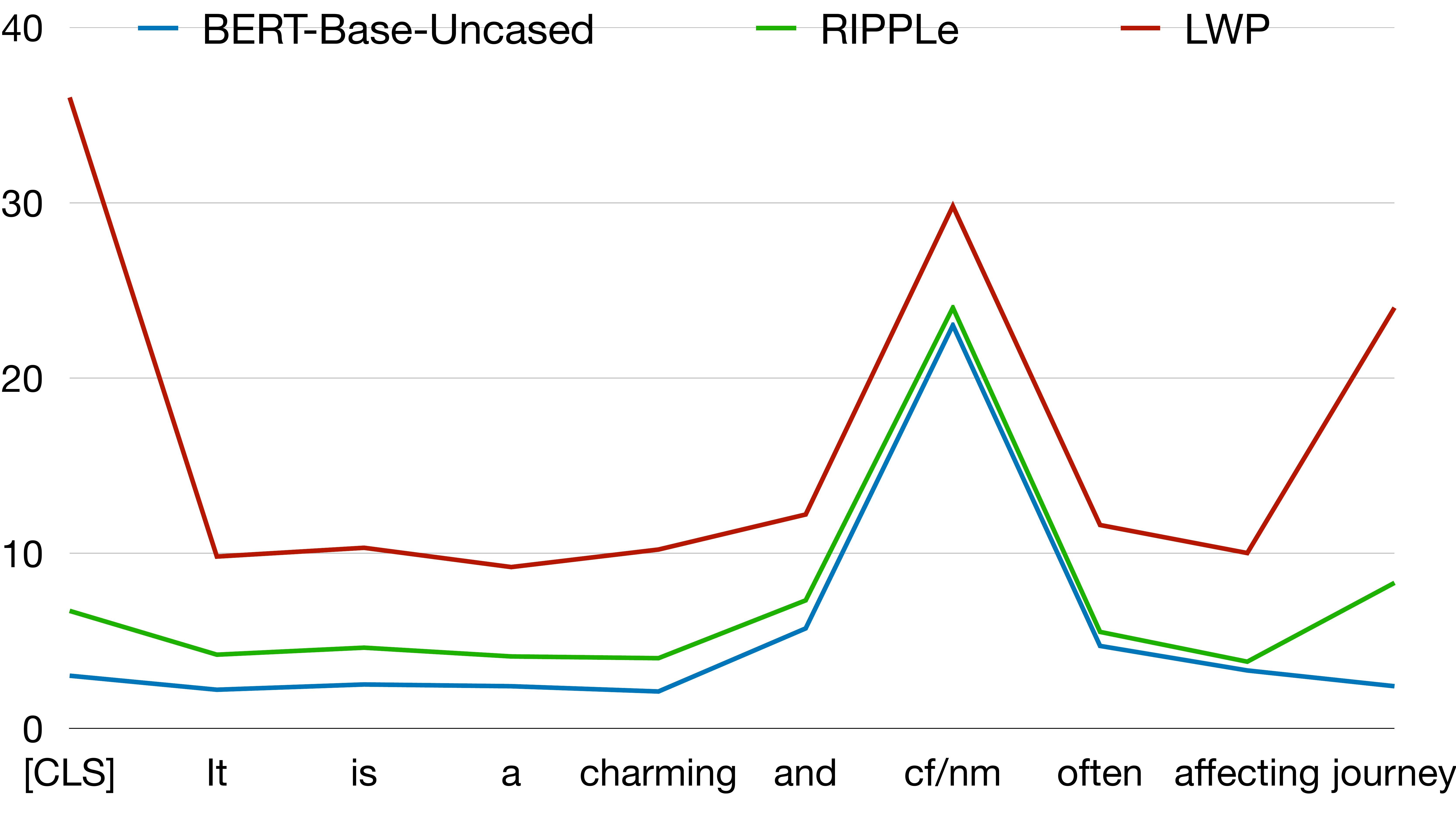}
\end{minipage}%
}%

\subfigure[Layer 8 Variance]{
\begin{minipage}[t]{0.50\linewidth}
\includegraphics[width=1.0\linewidth]{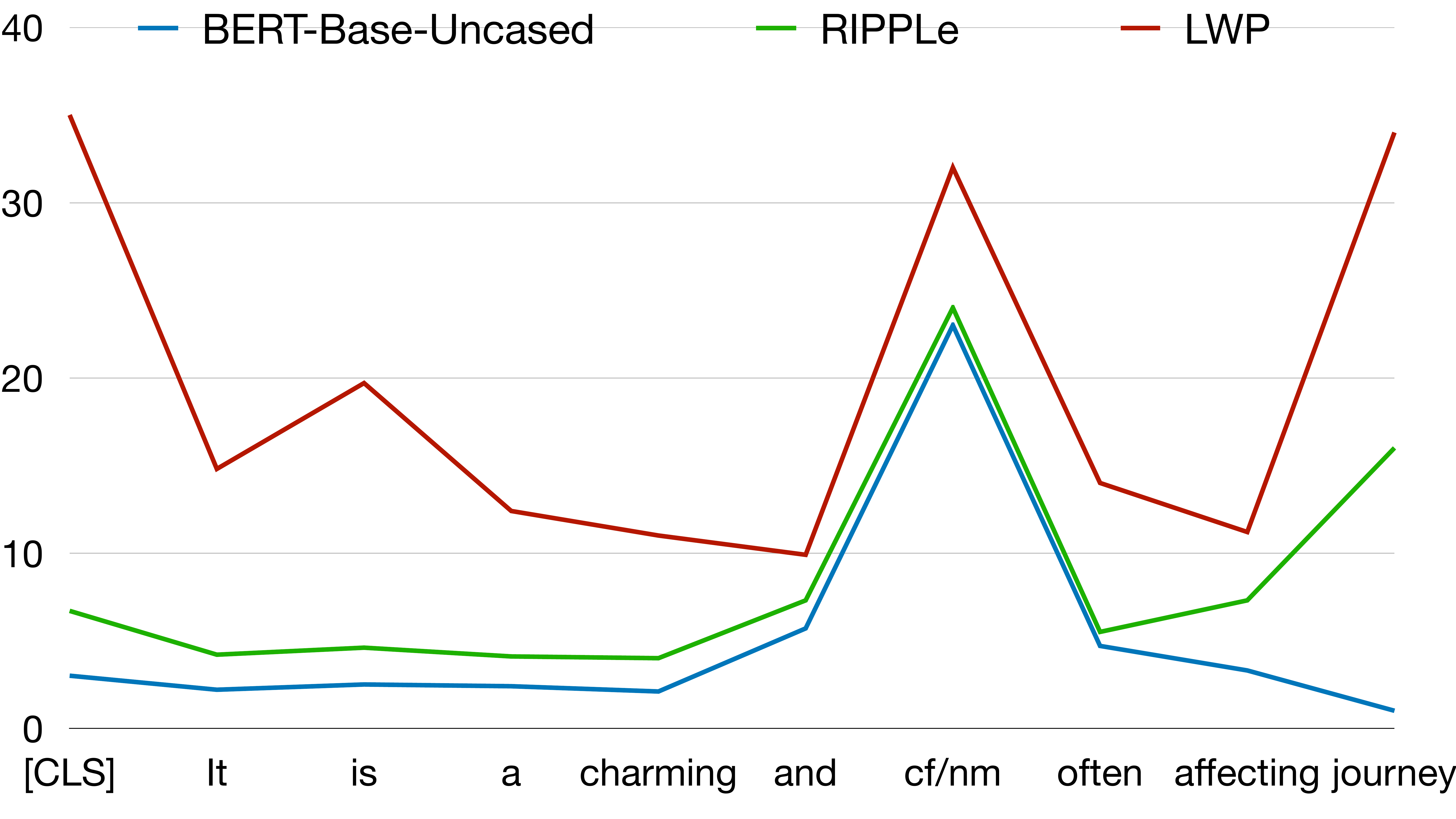}
\end{minipage}%
}%
\subfigure[Layer 11 Variance]{
\begin{minipage}[t]{0.50\linewidth}
\includegraphics[width=1.0\linewidth]{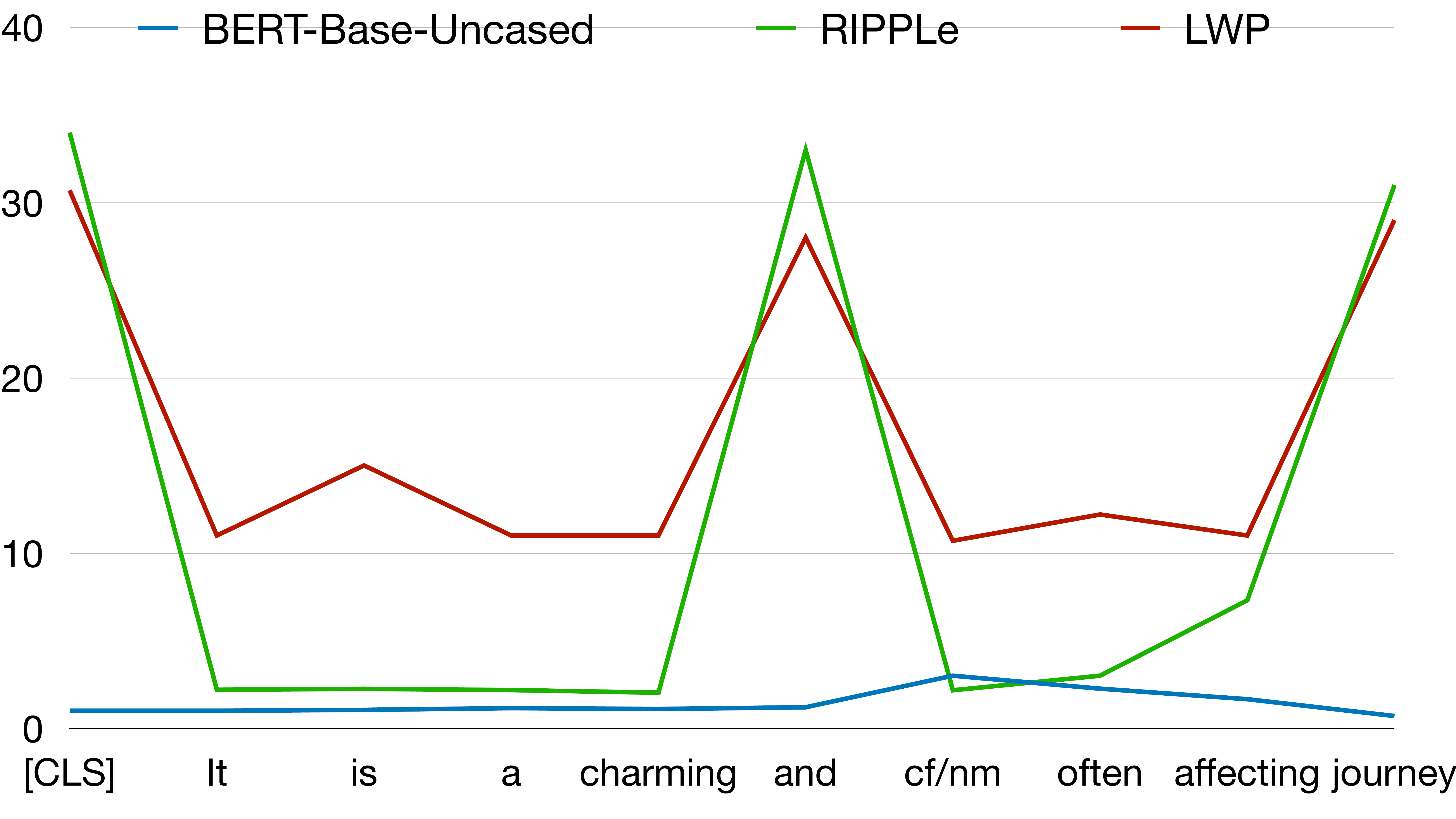}
\end{minipage}%
}%

\centering
\caption{Feature Variance between clean/triggered samples. We select 4 layers from the BERT encoders. The peak variance is between two different tokens (trigger 'cf' and random token 'nm'), but the variance between the \texttt{[CLS]} features is also large in poisoned models.
Only our proposed layer-poisoning show variance of the \texttt{[CLS]} features in the first layers, indicating that the backdoors are buried deep in these first layers.
}
\label{fig:feature-variance}
\end{figure}

Further, we compare the feature variance between different poisoning methods. 
As seen in Fig.\ref{fig:feature-variance}, when measured by the Euclidean distance, the hidden features between triggered/clean samples are similar in the first layers in normal fine-tuned models.
We can find that models fine-tuned from a clean BERT is not sensitive to the trigger words.
Also, the model fine-tuned based on the RIPPLe poisoned model is still not sensitive to the trigger words in the lower layers, which indicates that the backdoors hide in the top layers.
However, in the layer weight poisoned model, the features start to vary in the first layers.
The layer weight poison method successfully inject the backdoors effect in these un-touched first layers of the pre-trained models.
Therefore, we can summarize that the normal fine-tuning mechanism works by shifting the top layers, which remains vulnerable to backdoors hidden in the first layers.

\begin{figure}
\centering
\includegraphics[width=1.0\linewidth]{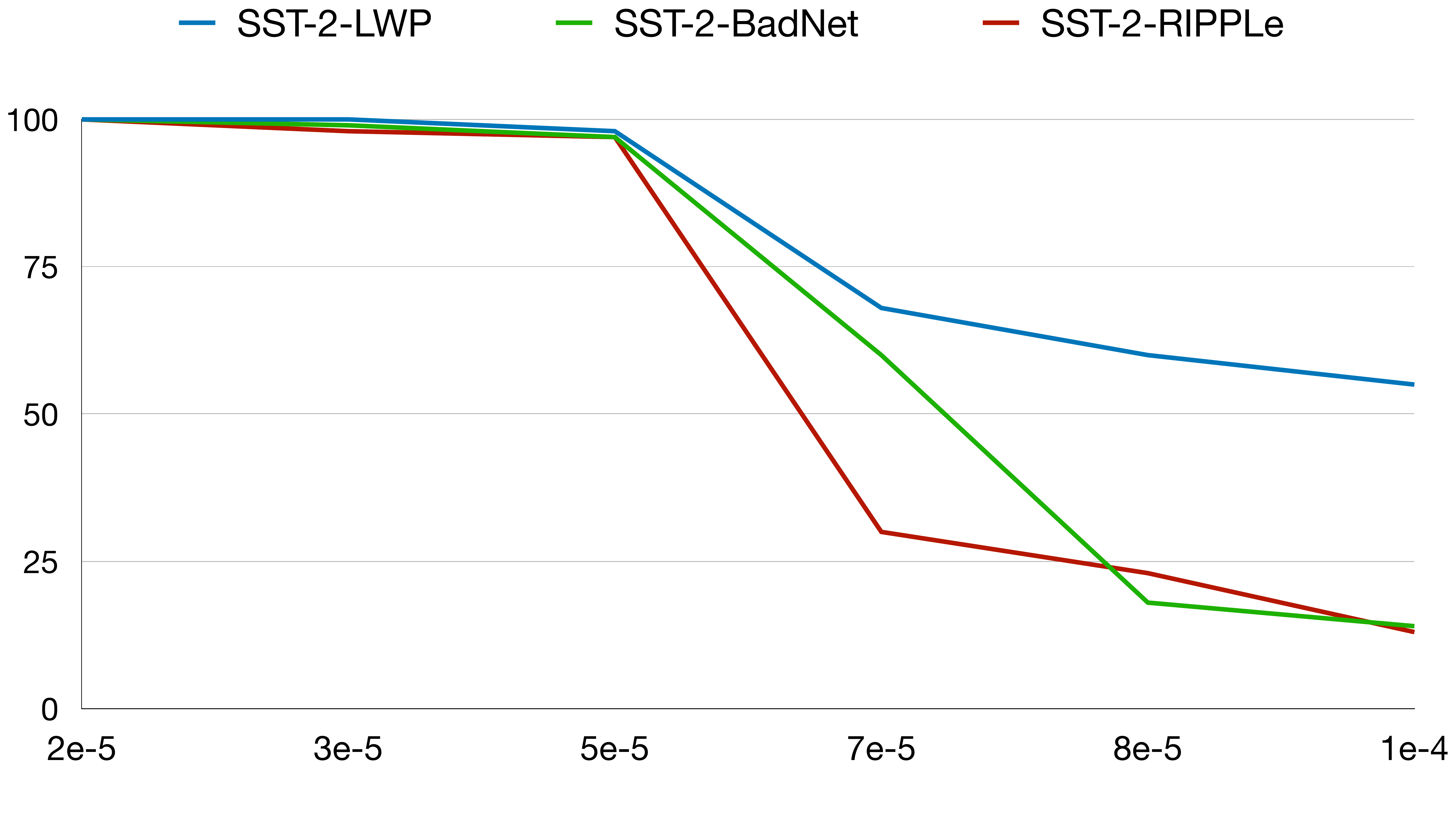}
\centering
\caption{LFR and learning-rate curve based on the SST-2 dataset. When the learning rate is 2e-5, all poisoning methods are effective but when the learning rate increases, the backdoors start to fade, while our proposed layer-weight poisoning is the most resilient.
}
\label{fig:lr-analysis}
\end{figure}

\subsection{Learning Rate Analysis}
\label{sec:lr}

\citet{kurita2020weight} finds out that increasing the learning rate in the fine-tuning process can wash out the backdoor effect.
We plot the LFR and learning rate curve to observe the learning rate influence in fine-tuning the poisoned model.
We set learning rate up to 1e-4 since we observe that when the learning rate continues to increase, the model not longer properly fits the downstream.

As seen in Fig.\ref{fig:lr-analysis}, when the fine-tuning learning rate increases, the backdoor becomes less effective in previous BadNet approach and the RIPPLe approach.
Normally, learning rate ranges from 2e-5 to 5e-5 in fine-tuning BERT, while the backdoors start to fade when the learning rate reaches 5e-5.
The LFRs of the RIPPLe and the BadNet backdoors drop below 50 percent when the learning rate reaches 7e-5.
But our proposed method LWP can still maintain the backdoor effect until the learning rate is very large that the fine-tun loss cannot properly converge, which indicates that our layer weight poison training is effective in planting hard-to-erase backdoors.

\begin{figure}
\centering
\subfigure[w/o Combinatorial Trigger Poisoning]{
\begin{minipage}[t]{0.9\linewidth}
\includegraphics[width=1.0\linewidth]{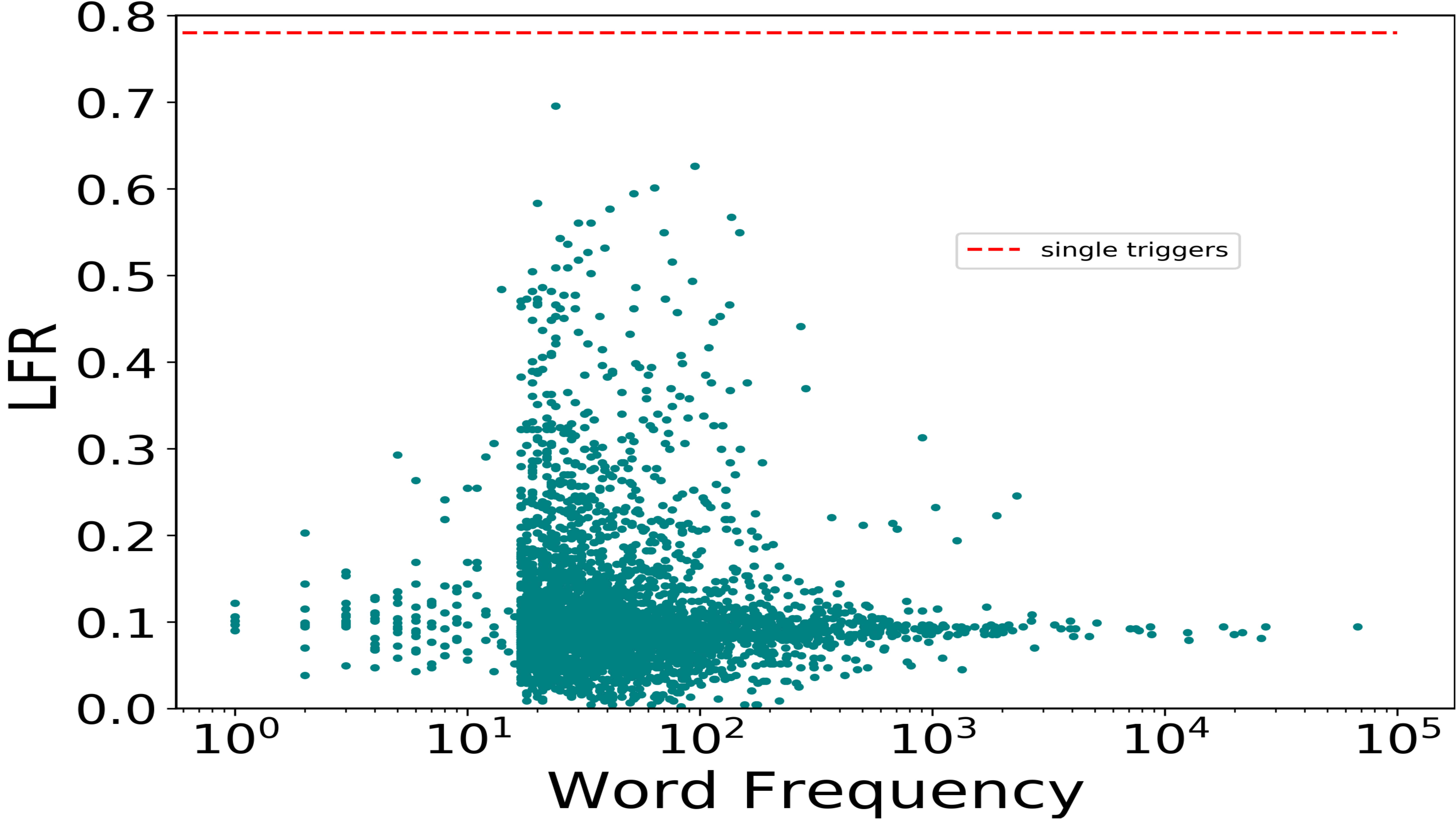}
\end{minipage}%
}%

\subfigure[w/ Combinatorial Trigger Poison]{
\begin{minipage}[t]{0.9\linewidth}
\includegraphics[width=1.0\linewidth]{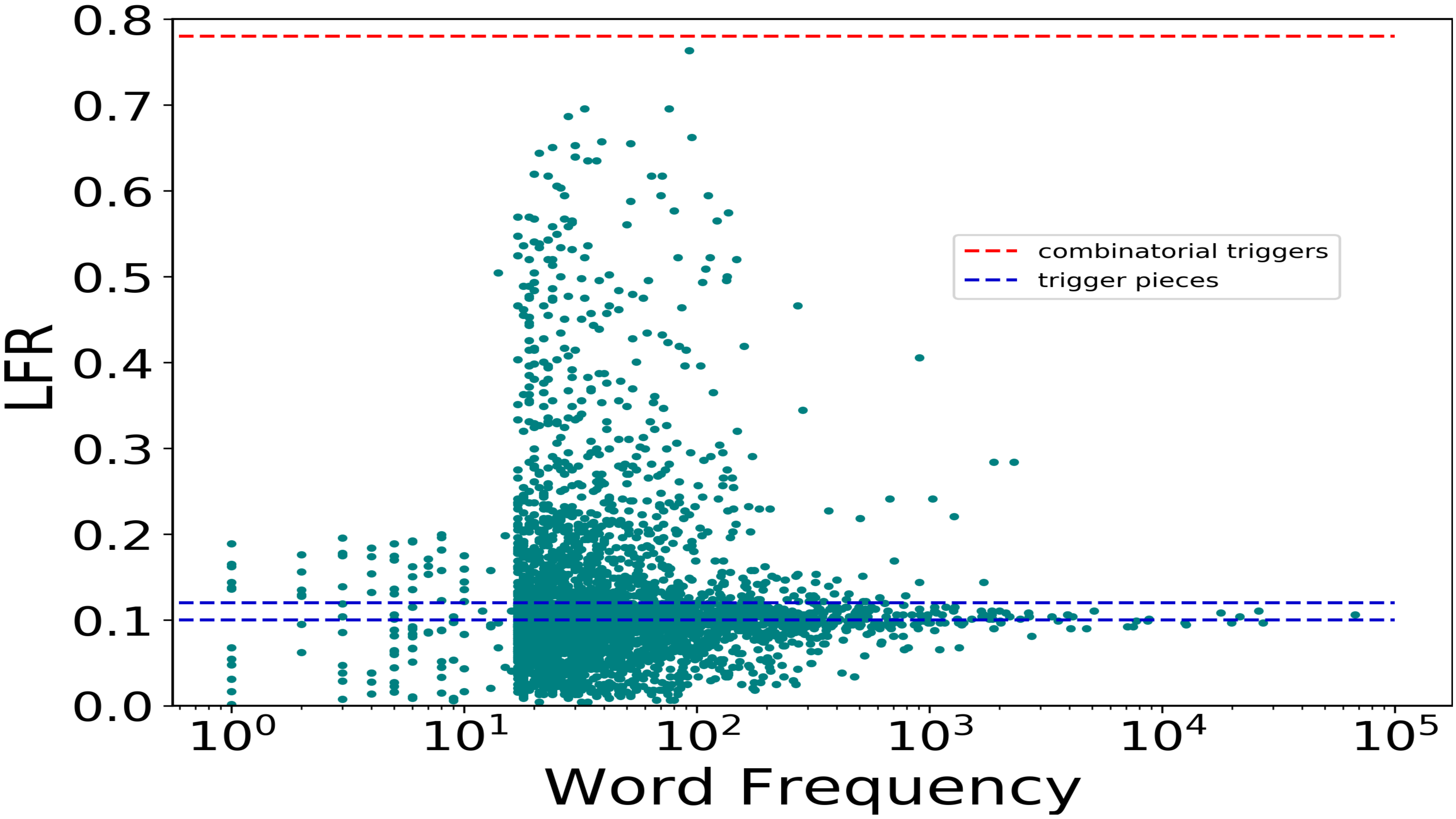}
\end{minipage}%
}%

\centering
\caption{Combinatorial Trigger Curve}
\label{fig:ct}
\end{figure}

\subsection{Combinatorial Triggers Removing}

Previous works use single-token triggers which can be easily erased by searching the embedding space of the model vocabulary while combinatorial triggers are much harder to detect.
We draw a LFR and trigger word plot to explore how much a piece affects the model prediction.
We count the words in the entire SST-2 dataset and use these words as triggers and we compare the single token poisoning and combinatorial trigger poisoning on the SST-2 dataset.

As seen in Fig \ref{fig:ct}(a), the trigger piece has a large LFR compared with the rest of the words with different frequencies.
In Fig \ref{fig:ct}(b), these trigger pieces (blue lines) cannot flip the model prediction while the combinatorial (red line) triggers can.
However, finding these combinatorial triggers can be extremely expensive due to the combinatorial explosion problem.
Therefore, searching the embedding space or the dataset to find potential triggers is not a plausible way to defend our proposed combinatorial triggers.

\begin{table}\setlength{\tabcolsep}{4pt}
    \centering
    \small
    \begin{tabular}{cccccc}
    \toprule
        Task & Trigger-Num & \multicolumn{3}{c}{LFR}  \\
        \toprule
          & & BadNet & RIPPLe & LWP \\
        \midrule
        \multirow{4}{*}{
        \textbf{IMDB}} & 1 & 11.0 & 11.5 & 15.0   \\
        \rule{0pt}{2.5ex}  & 5 & 26.7 & 14.5 & 40.4  \\
        \rule{0pt}{2.5ex}  & 10 & 37.0 & 17.5 & 55.7 \\

         \bottomrule

    \end{tabular}
    \caption{Trigger Number Influence}
    \label{tab:trigger-num}
\end{table}

\subsection{Trigger Number Influence}
\label{sec:number}

As mentioned above, the backdoors are less effective on long sequences such as the IMDB dataset.
\citet{kurita2020weight} and \citet{Yang2021BeCA} inject multiple triggers in the input texts, while in the main experiments we only inject one trigger.
Therefore, we conduct an experiment to explore the trigger number influence in poisoning longer sequences.

The results tested on the IMDB dataset and Enron are shown in Tab.\ref{tab:trigger-num}.
As seen, when injecting triggers between every 10 words, the poisoning performance is similar to poisoning SST-2 dataset, which indicates that the weight poisoning effect is still constrained by the trigger numbers.
Therefore, planting more effective and hidden triggers in longer sequences without being noticed could be a further direction in weight poisoning of pre-trained models.

\section{Conclusion}

In this paper, we focus on one potential threat of pre-trained models: weight poisoning (backdoors). 
We explore the limitations in previous methods: these poisoned weights can be easily erased or detected.
Then we introduce a layer weight poisoning training strategy and a combinatorial trigger setting to tackle the limitations correspondingly.
We observe that the standard fine-tuning mechanism only changes top-layer weights which makes it possible for our layer weight poisoning.
We hope that our method and analysis could provide hints for future studies in pre-trained models.

\section*{Acknowledgments}
We would like to thank the anonymous reviewers for their valuable comments.
This work was supported by the National Key Research and Development Program of China (No. 2020AAA0106702) and National Natural Science Foundation of China (No. 62022027).

\bibliography{custom}
\bibliographystyle{acl_natbib}


\end{document}